\documentclass[12pt]{article}
\usepackage{color,multicol}
\usepackage{amsmath}
\usepackage{amssymb}
\usepackage{epsfig}
\usepackage{pifont}

\usepackage{tocloft}

\begin{document}

\title{ECONOMICS OF DISAGREEMENT\\ --\\ financial intuition for the R\'enyi divergence\footnote{Original version 18 November 2018. {\sl Journal-ref:} Entropy {\bf 22}(8), 860 (2020).\vspace*{1mm}}
}
\author{Andrei N. Soklakov\footnote{
Strategic Development, Deutsche Bank; Andrei.Soklakov@(db.com, gmail.com)\vspace*{1mm}\newline
{\sl The author is grateful to Michael Shadlen for his insightful comments and helpful consultations on the neurophysiology of decision making.
The views expressed herein should not be considered as investment advice or promotion. They represent personal research of the author and do not necessarily reflect the view of his employers, or their associates or affiliates.} }}
\date{}
\maketitle

\begin{center}
\parbox{14.4cm}{ 
{\small
Disagreement is an essential element of science and life in general. The language of probabilities and statistics is often used to describe disagreements quantitatively. In practice, however, we want much more than that. We want disagreements to be resolved. This leaves us with a substantial knowledge gap which is often perceived as a lack of practical intuition regarding probabilistic and statistical concepts.\\

Take for instance the R\'enyi divergence which is a well-known statistical quantity specifically designed as a measure of disagreement between probabilistic models. Despite its widespread use in science and engineering, the R\'enyi divergence remains a highly abstract axiomatically-motivated measure. Certainly, it offers no practical insight as to how disagreements can be resolved.\\

Here we propose to address disagreements using the methods of financial economics. In particular, we show how a large class of disagreements can be transformed into investment opportunities. The expected financial performance of such investments quantifies the amount of disagreement in a tangible way. This provides intuition for statistical concepts such as the R\'enyi divergence which becomes connected to the financial performance of optimized investments. Investment optimization takes into account individual opinions as well as attitudes towards risk. The result is a market-like social mechanism by which funds flow naturally to support a more accurate view. Such social mechanisms can help us with difficult disagreements (e.g., financial arguments concerning the future climate). \\

In terms of scientific validation, we used the findings of independent neurophysiological experiments as well as our own research on the equity premium.\\

\phantom{{\sl Keywords:} }risk aversion; information derivatives; neuroeconomics.\\

}
}
\end{center}


\section{Introduction}

In a lot of practical situations knowledge about the future can be expressed in terms of possible events and their probabilities. In such situations disagreements can be understood as a mismatch between probability distributions (often advocated by different people).

The extent of disagreement between probability distributions is routinely measured by specially designed mathematical quantities called divergences. Relative entropy~\cite{KullbackLeibler_1951} and its celebrated generalization -- the R\'enyi divergence~\cite{Renyi_1961} -- are among the best known examples of such quantities.

Unfortunately, the numerical values of abstract divergences can be difficult to interpret (see below for an example). Moreover, the standard axiomatic definitions of divergencies contain no pragmatic insight as to how one can resolve the underlying disagreements in practice.

Here we revisit one of the oldest ways of resolving disagreements -- by settling bets. The financial performance of a well-designed bet reflects the extent of the disagreement in an intuitive way. The betting mechanism itself provides a practical algorithm for resolving the disagreement.

On the mathematical side, our work is closely related to Kelly's financial interpretation of relative entropy~\cite{Kelly_1956}. Our interpretation can be used for the R\'enyi divergence (which includes Kelly's main result as a special case) and can be easily generalized to other more sophisticated divergence measures. On the scientific side, we ensure the financial interpretation is used within carefully tested limits. This is done by checking against the observed economic data and against independent results in experimental neurophysiology.

Readers familiar with Kelly's work and the concept of divergencies may proceed to the next section. The rest of this introduction is mostly pedagogical in nature. Here we assume minimal prior knowledge and attempt to broaden the audience by providing additional background material.\\
    \centerline{\rule[1ex]{.25\textwidth}{.5pt}}
The lack of accurate intuition has always been a major challenge in the fields of probability and statistics. Yet, transcending the centuries, one particular technique appears time and time again as a popular source of statistical intuition. Found in virtually every course on probability and statistics, the technique involves imagining a game of chance (throwing dice, tossing coins, playing cards and lotteries) with well-defined financial outcomes.

In 1956 Kelly used this technique to propose an apparently very general and intuitive interpretation of relative entropy~\cite{Kelly_1956}. Kelly considered a growth-optimizing investor in a game with mutually exclusive outcomes (a \lq\lq horse race'') and showed that the rate of return expected by such an investor is equal to the relative entropy between the investor's believed probabilities and the official odds.

Effectively, Kelly showed that a growth-optimizing investor would expect on average to convert 1 bit of additional information (relative entropy) into 100\% in financial returns.

Unfortunately, this rather spectacular conversion of information into financial returns ran into conflict with the ideology of efficient markets which dominated mainstream economics at the time. Efficient markets were believed to be unbeatable in their ability to include all information. This left no logical room for any investor to disagree with the market. Any such investor was deemed to be making a fatal error of judgement before they could even present their reasoning.

After a decade of obscurity Kelly's ideas began to bypass mainstream academia and, with support from independent scholars, started to influence financial decisions at the applied level of fund managers (see~\cite{MacLeanEtAl_2011} and references therein). This attracted vigorous criticism from the very top of the academic establishment. Probably the most influential criticism came from Samuelson, who received the Nobel Memorial Prize in Economics in large part due to his highly influential work on the efficient market hypothesis.

Kelly's mathematics was sound, so instead, Samuelson focused his critique on possible misunderstandings~\cite{Samuelson_1971,Samuelson_1979}. In our context of developing strong statistical intuition such critiques can be very useful. However, we should also exercise a great deal of caution to avoid propagation of bias. Readers unfamiliar with academic economics might better appreciate the challenge by inspecting the writing style of Ref.~\cite{Samuelson_1979} cited above. Ref.~\cite{Samuelson_1979} was painstakingly edited to contain exclusively monosyllabic words. Neither the referees nor the editors could question Samuelson's authority and one can still find prominent academics citing the monosyllabic presentation as evidence of sharp intellect.

One way of purifying Samuelson's critique to a scientific standard is to state it as an observation that real people are not necessarily growth-optimizing. This observation is testable and, indeed, undoubtedly true. If that is so, real people might be misled by the information-theoretic intuition designed by Kelly for the growth-optimizing investor. At the very least we must agree to investigate the robustness of the intuition with respect to more realistic investors.

Here we show that the connection between information and real-life expected financial returns is in fact very robust. To this end we consider investors with different levels of risk aversion, $R$. The maximum expected returns coincide with the relative entropy. This is the case of the growth-optimizing investor ($R=1$) considered by Kelly. Any deviation from $R=1$ causes a drop in expected returns. However, the amount of the drop is also information-driven: in the case of constant relative risk aversion, for example, it is proportional to the absolute difference between the relative entropy and the R\'enyi divergence.

In other words, attempts to challenge the financial intuition with more general investment behaviors simply upgrade it for a wider class of divergence measures.

However, because real investors are not completely general (mathematically speaking), there are scientific limits beyond which our financial intuition should not be applied. We show where these limits are in the case of the R\'enyi divergence.

The financial interpretation of the R\'enyi divergence turns out to be remarkably intuitive. We illustrate this point in the next section by giving a numerical example. Here we just mention that human intuition for financial returns is indeed quite accurate. A quick look at the savings rates offered by different banks (through popular price comparison websites) shows some financial returns specified to the accuracy of $0.01\%$ per annum. Similar accuracy can be seen on mortgage rates and on foreign exchange commissions.

Even if we round this up by two orders of magnitude to 1\% we still get a lot of accuracy. Intuitive appreciation of abstract statistical quantities such as the R\'enyi divergence with this kind of numerical accuracy is a remarkable prospect.

The curious reader may want to review our intuition for the R\'enyi divergence against alternatives. Unfortunately, despite the widespread use of the R\'enyi divergence in science and engineering, we could not find alternatives with similar levels of generality or intuitive accuracy. The best examples we could find require fairly specialized knowledge. In coding theory, for example, the R\'enyi divergence can be understood as the possible amount of compression in a probabilistic mixture of two codes~\cite{Harremoes_2006,Gruenwald_2007}. In decision theory the R\'enyi divergence can be linked to error probability. For example, in a two-sensor composite hypothesis testing framework the R\'enyi divergence can be interpreted as the optimal worst-case miss-detection exponent~\cite{Shayevitz_2011}. Such observations are undoubtedly very useful within their respective fields yet difficult to adopt across a wide range of disciplines.

In contrast, we expect the financial approach to become a general purpose technique covering a broad range of applications. It can even be used as an organizational instrument for resolving disagreements, raising financial support and accelerating the adoption of sound scientific developments. This is because financial returns can be realized and used in practice.

The intuitiveness of the financial approach is an interesting phenomenon in its own right. Somehow our brains are much better at understanding financial returns than divergencies despite a simple mathematical connection. It is not unreasonable to expect that such highly intuitive techniques might have some universal physiological basis. Inspired by this proposition as well as by the standard requirement of on-going scientific testing we searched for independent neurophysiological studies of decision making. We found a similarity between the financial returns discussed in this paper and the firing rates of certain neurons implicated in decision making~\cite{YangShadlen_2007,KiraEtAl_2015}.
\begin{center}
    \rule[1ex]{.25\textwidth}{.5pt}
\end{center}
The rest of the paper is organized as follows. In section 2 we present the results by giving a practical illustration. Section 3 concludes the presentation of the results with a short discussion. All technical details are deferred to section 4. This has four parts. Subsection 4.1 provides the bulk of mathematical derivations. Subsection 4.2 uses economic data on the equity premium to determine the scientific bounds of the financial intuition. In subsection 4.3 we demonstrate consistency of the financial intuition with the recent neurophysiological discoveries. Finally, subsection 4.4 discusses the possibility of further generalizations.

\section{Results}\label{Sec:Illustration}

We promised the reader an intuitive economic technique for resolving disagreements (with a built-in intuition for the R\'enyi divergence). To see how this works, imagine an argument between two scientists, Maggie and Bob, about the distribution of outcomes for some experiment $X$. Maggie believes $m$ is the correct distribution while Bob thinks the real distribution is $b$. The R\'enyi divergence of order $\alpha$ of $b$ from $m$ reads~\cite{Renyi_1961}
\begin{equation}\label{Eq:D_alpha}
D_\alpha(b||m)\overset{\rm def}{=}\frac{1}{\alpha-1}\ln\int b(x)\Big(\frac{b(x)}{m(x)}\Big)^{\alpha-1}\,dx\,,
\end{equation}
where the integration is performed over all possible outcomes of the experiment $X$.

\begin{figure}[t!]
\includegraphics[width=\textwidth]{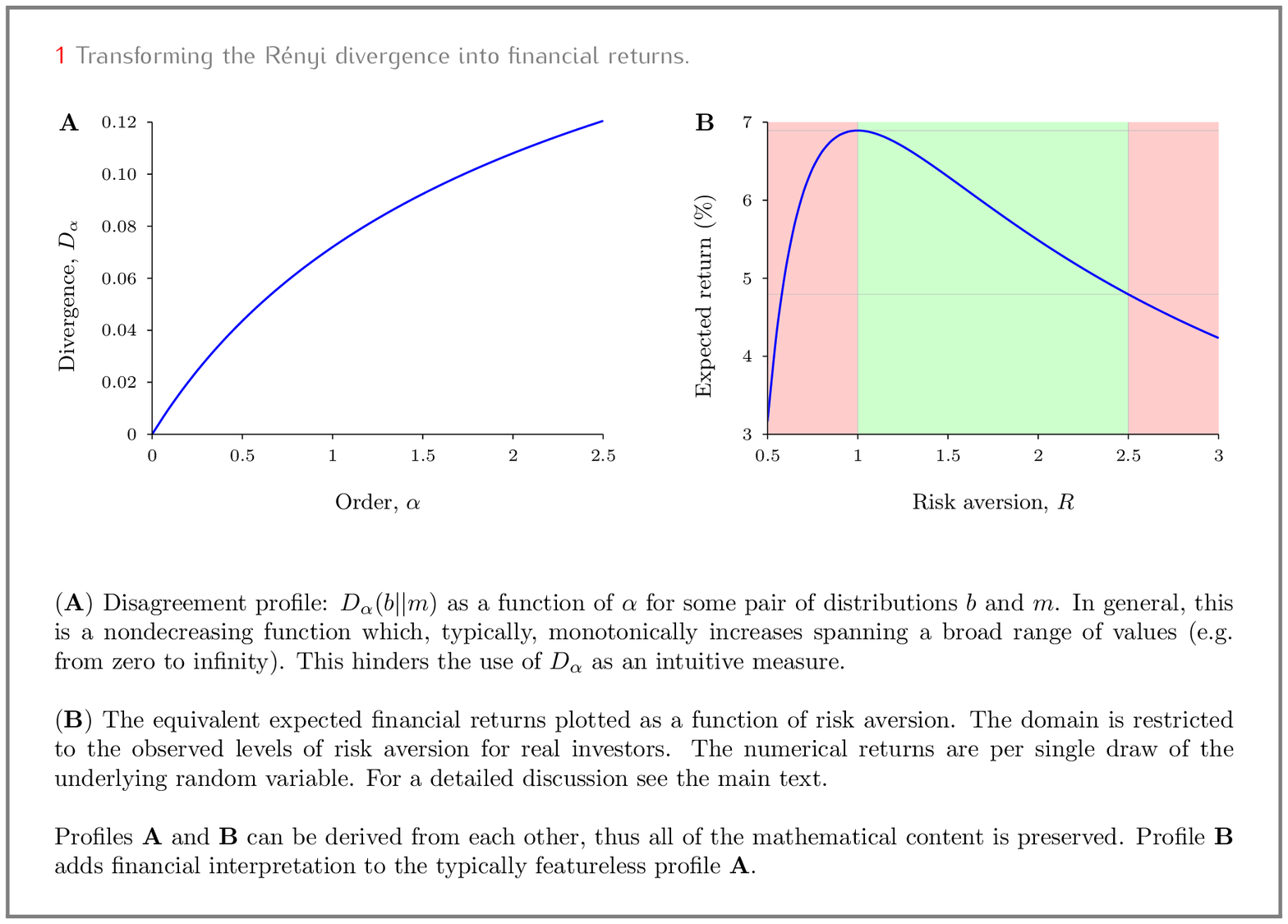}\label{Fig:RenyiReturn}\\
\end{figure}

Figure~1A provides a concrete numerical example of a possible disagreement profile ($D_\alpha$ as a function of $\alpha$). We want to understand intuitively the amount of disagreement in Fig.~1A (using nothing but the data in Fig.~1A). Because we want the R\'enyi divergence to speak for itself, we deliberately refrain from providing any further context. Indeed, the whole point of using divergence measures is to understand the overall amount of disagreement in a way that is meaningful in its own right, i.e. without having to inspect the individual conflicting views in detail.

Even though the R\'enyi divergence was specifically designed as a measure of disagreement between distributions, Fig.~1A provides the two scientists with little practical intuition on the extent of their disagreement (not to mention any idea on how it can be resolved).

Maggie happens to be very rich so she decides to prove her point by offering a fair price for any game regarding the experiment. The game is specified by its payoff function $F$ which states the amount of reward, $F(x)$, associated with a particular outcome, $x$, of the experiment. For any such game Maggie computes
\begin{equation}\label{Eq:Price}
{\rm Price}[F]\overset{\rm def}{=}\int F(x)\,m(x)\,dx\,.
\end{equation}
Maggie's decision to provide tradable prices in the above manner is a standard service in the financial industry known as \lq\lq market making''.\\

This gives Bob a chance to defend his view by demonstrating a sizeable profit. Indeed, if Maggie's estimation of probabilities was not accurate her pricing would be wrong and Bob could try to choose the shape of $F$ so as to benefit from the mispricing.

Bob calculates that the right choice of $F$ allows him to grow his capital exponentially. For the expected growth rate Bob computes (see section~\ref{Sec:Derivations} for details)
\begin{equation}\label{Eq:RExpectedLogReturn}
{\rm ExpectedRate}=\frac{1}{R} D_{1}(b||m)+\frac{R-1}{R}D_{1/R}(b||m)\,,
\end{equation}
where $R$ is Bob's personal level of risk aversion. The definition for $R$ is given in section~\ref{Sec:Derivations}. From a practical standpoint it is important to know the realistic range of $R$ describing real people. In section~\ref{Sec:RiskAversionRange} we use economic data and find evidence of $R$ varying between 1 and 2.5. For the purposes of this paper we refer to this range of risk aversion as natural and refrain from making any statements outside this range (despite much wider mathematical validity).

Bob uses Eq.~(\ref{Eq:RExpectedLogReturn}) to transform the non-intuitive profile of Fig.~1A into the expected rate of return as a function of his risk aversion (see Fig.~1B). The benefits of this transformation are immediate; Bob understands the extent of his disagreement with Maggie. Specifically, he expects to make on average 4.8\%-6.9\% per run of the experiment\footnote{The actual returns are random but, in the long run, Bob can hope to achieve the expected growth of Fig.~1B almost surely by simple repetition, i.e.\ by reinvesting the actual returns from one run of the game into the next. This is, of course, assuming that Bob's belief $b$ turns out to be correct. In general, the long-term realized returns will follow an instance of Eq.~(\ref{Eq:GeneralExpectedLogReturn}) where $p$ is chosen to coincide with the correct (i.e.\ the realized) distribution.} (without taking uncomfortable risks, i.e. staying within the natural range of risk aversion).

Aiding Bob's understanding there are many real-life economic benchmarks. At the household level he can compare his game with Maggie with the interest on his savings or loans. If he prefers the macro level, he can use figures such as GDP growth or inflation.

As we pointed out in the introduction, the accuracy with which financial and economic benchmarks are published tells us that people are quite sensitive to financial returns. This natural sensitivity can be further amplified through compounding. Given a statistically significant set of 100 runs, for example, Bob would expect his capital to grow on average 120 to 986 times. The enormity of these expected returns should be obvious to anyone (even to people who have never had a bank account).

To implement his strategy Bob would of course need to know his own view $b$ as well as the market $m$. However, it is worth emphasizing that the above materiality analysis does not depend on such detailed knowledge. To comprehensively understand the overall scale of the disagreement the only information we might want to know in addition to the disagreement profile (Fig.~1A) is information about the actual resources committed by the players including how often (or how many times) the game can be played in practice.

In the above example Maggie took on the extra responsibility of making the market. This was useful to simplify the illustration. In the most general practical case we might be dealing with multiple players each with their own believed distribution, risk aversion and a limited budget, and there might not be anybody among the players brave enough to make the market. In this case the market should form spontaneously (assuming, of course, everyone can see the opportunity). The corresponding distribution $m$ would be computed as explained in Appendix V of the supplementary materials paper~\cite{Soklakov_2014EqPuzzle}. For example, a group of growth-optimizing investors with views $\{b_i(x)\}$ and budgets $\{w_i\}$ will form a market with $m(x)=\sum_i w_ib_i(x)/\sum_k w_k$. Whether the market forms spontaneously or is made by one of the players, equations (\ref{Eq:RExpectedLogReturn}) and (\ref{Eq:GeneralExpectedLogReturn}) hold for every individual investor.

The above illustration appeals to the apparent fact that people find it easier to make decisions in terms of financial returns rather than the specially designed entropic quantities such as the R\'enyi divergence. Some readers may be happy to take this fact for granted (as an obvious empirical observation). Others are invited on a small detour of section~\ref{Sec:Neuroscience} where we open the study of the curious intuitiveness of financial returns as a cognitive phenomenon.

As a bonus, let us mention that the financial approach also aids structural understanding of disagreements. This can be gained by inspecting the payoff functions (in addition to the expected returns considered in this paper). The relevant discussion would take us deeper into finance and require some familiarity with financial products. The interested reader is referred to Ref.~\cite{Soklakov_2014MR} for further details.

\section{Discussion}

Throughout history, mankind's ability to plan and optimize was key to our survival. This gave us economic intuition which can be surprisingly strong. Here we demonstrated its power to resolve complex (statistical) disagreements. The concluding comments below briefly touch on further generalizations and practical applications.

On a purely mathematical level, further generalization of our results is almost too easy. More complex investment behaviors naturally lead to a wider class of divergence measures (see section~\ref{Sec:MathGeneralizations}). Such ease calls for scientific caution. Any generalization (as well as our current understanding) should be kept in sync with experimental findings.

Regarding practical applications, we have already demonstrated that framing statistical disagreements in terms of financial optimization quantifies the amount of disagreement in a way that is both intuitive and  scientifically sound (as far as the current experimental knowledge suggests). This gives us confidence to recommend our approach as a mechanism for closing real disagreements in a cooperative manner.

It would be very interesting, for instance, to see what financial returns modern climatology can demonstrate in a game against sceptics of climate change. We can set up such a game just as we did above for Maggie and Bob. All we would need are concrete examples of some relevant probability distributions (preferably with short time horizons to facilitate convergence via repetition).

The financial attractiveness of the optimized investments is directly related to the amount of genuine disagreement between participants. The optimization ensures the disagreement is utilized to its full potential (from the point of view of each individual participant). Following the optimization, failure to trade would imply that either the disagreement is not large enough (relative to other investment opportunities) or the disagreement itself is not genuine (i.e. it fails to capture the real problem).

\section{Materials and Methods}

\subsection{Derivations}\label{Sec:Derivations}

A game of chance is defined by a random variable, $X$, and a payoff function, $F$. The value $F(x)$ states the amount of reward associated with a particular outcome, $x$, of the variable. This technical definition covers many practical situations. For instance, $F(x)$ might denote the payout for medical insurance in the event of a diagnosis $x$.

Consider an investor who is interested in finding the optimal shape of $F$ so as to maximize their benefit. Following von~Neumann and Morgenstern~\cite{NeumannMorgenstern_1944} we understand the benefit as the expected utility of the payoff
\begin{equation}\label{Eq:ExpectedUtility}
E_bU[F]\overset{\rm def}{=}\int b(x)\, U(F(x))\,dx\,,
\end{equation}
where $U$ is the utility function and $b$ is the investor-believed probability distribution for the variable $X$. It is important to remember that both the probability distribution $b$ and the utility function $U$ reflect the investor's personal understanding of the game.

When maximizing $E_b U$ the investor is subject to a budget constraint. To state this constraint mathematically we need the ability to price the game. Assuming a very general setting which is used in the financial industry we write the fair price of $F$ as the average~(\ref{Eq:Price}), where the positive-valued function $m$ summarizes the relative prices of ensuring against every outcome $x$ of the random variable $X$. In game-theoretic illustrations this is often stated in terms of \lq\lq odds''; economists would recognize this as Arrow-Debreu prices. For our purposes, the important point to remember is that $m$ is a given property of the market and the investor has no choice but to take it into account when understanding their budget. For simplicity we assume that $m$ is a probability distribution (\lq\lq fair odds'') and that ${\rm Price}[F]=1$.

The payoff $F$ which achieves the maximum of $E_b U[F]$ under the constraint ${\rm Price}[F]=1$ satisfies the payoff elasticity equation~\cite{Soklakov_2013b}
\begin{equation}\label{Eq:PayoffElasticity}
\frac{d\,\ln F}{d\,\ln f}=\frac{1}{R}\,,
\end{equation}
where $f=b/m$ and $R=-FU''_{FF}/U'_F$ is the Arrow-Pratt relative risk aversion.
This equation is used in finance to produce {\it information derivatives} -- financial instruments which are derived from all relevant information (including market-implied $m$ together with investor-believed $b$ and $R$).

Kelly's intuition concerns the important special case of $R=1$. This is the famous case of the growth-optimizing investor which was first introduced by Bernoulli in 1738~\cite{Bernoulli_1738}. Indeed, when $R=1$ the utility function $U(F)\propto\ln(F/{\rm Price}[F]) + const$, so the investor is optimizing the expected logarithmic rate of return:
\begin{equation}\label{Def:ExpectedLogReturn}
E_b\,{\rm Rate}[F]=\int b(x)\,{\rm Rate}(F(x))\,dx\,,\ \ \  {\rm Rate}(F(x))\overset{\rm def}{=}\ln\Big(F(x)/{\rm Price}[F]\Big)\,.
\end{equation}

Kelly studied the growth-optimizing investor in detail and showed that under some natural assumptions the investor's expectation for the logarithmic rate of return is exactly the relative entropy~\cite{Kelly_1956}. The reader can verify this result independently by substituting $R=1$ into Eq.~(\ref{Eq:PayoffElasticity}) and computing
\begin{eqnarray}\label{Eq:maxLogReturn}
\max_F E_b\,{\rm Rate}[F]&\underset{(R=1)}{=}&E_b\,{\rm Rate}[f]\cr
&=&\int b(x)\,\ln\frac{b(x)}{m(x)}\,dx\,.
\end{eqnarray}
This is the mathematical essence of Kelly's game-theoretic (financial) interpretation for the information rate.

Kelly's interpretation depends on the investor being growth-optimizing. According to Samuelson, this is a major weakness which limits (if not prevents) any use of Kelly's interpretation in practice~\cite{Samuelson_1971,Samuelson_1979}.

In an entirely separate argument, R\'enyi considered the core mathematical properties of information, formulated them as axioms, and recognized relative entropy (the last expression in Eq.~(\ref{Eq:maxLogReturn})) as a special case of a much larger class of information measures~\cite{Renyi_1961}. This class is spanned by linear combinations of the quantities defined in~(\ref{Eq:D_alpha}) with different values of $\alpha$. An individual $D_\alpha(b||m)$ is called R\'enyi's divergence of order $\alpha$ of a probability distribution $b$ from another distribution $m$. The relative entropy is included in this definition as the limiting case~\cite{Renyi_1961}
\begin{equation}\label{Eq:D_1}
D_1(b||m)\overset{\rm def}{=}\lim_{\alpha\to 1}D_\alpha(b||m)=\int b(x)\,\ln\frac{b(x)}{m(x)}\,dx\,.
\end{equation}

In what follows we show that Samuelson's demands for considering more general investors and R\'enyi's generalization of the relative entropy are in fact a statement and a solution of the same economic problem which translates information content (captured as disagreement between distributions) into financial returns.

Let us consider an investor with an arbitrary constant relative risk aversion, $R=const$. Unless $R=1$, i.e. unless the investor is growth-optimizing, the growth rate expected by the investor will be smaller than the Kelly benchmark~(\ref{Eq:maxLogReturn}). The natural question to ask is how much smaller. Using Eq.~(\ref{Eq:PayoffElasticity}) we derive the optimal payoff
\begin{equation}
F(x)=\frac{f^{1/R}(x)}{{\rm Price}[f^{1/R}]}\,.
\end{equation}
By direct substitution into Eq~(\ref{Def:ExpectedLogReturn}) we compute
\begin{equation}
E_b\,{\rm Rate}[F]=\frac{1}{R} E_b\,{\rm Rate}[f]+\frac{R-1}{R}D_{1/R}(b||m)\,.
\end{equation}
Together with Eqs.~(\ref{Eq:maxLogReturn}) and (\ref{Eq:D_1}) this gives us the transformation~(\ref{Eq:RExpectedLogReturn}). As a side observation, the reader might be interested to note that the structure of this expression is rather general. In particular, by replacing the investor-believed distribution $b$ with any other distribution $p$ one can prove a much more general law with the exact same structure. Expanding $E_p\,{\rm Rate}[f]=D_1(p||m)-D_1(p||b)$ one can write this as
\begin{equation}\label{Eq:GeneralExpectedLogReturn}
E_p\,{\rm Rate}[F]=\frac{1}{R} \Big( D_1(p||m)-D_1(p||b)\Big) +\frac{R-1}{R}D_{1/R}(b||m)\,.
\end{equation}
In other words, one can talk about the investor-expected returns ($p=b$) or look at the actual realized returns (in which case $p$ coincides with the actual distribution for $X$) or we can take the perspective of a totally independent observer who computes expectations using a very different distribution $p$ and, in all of these cases, the effect of risk aversion on the financial performance of the growth-optimizing investor would follow the same universal law~(\ref{Eq:GeneralExpectedLogReturn}).

Coming back to the investor-expected returns, let us investigate the drop in the expected return of $F$ relative to the growth-optimizing $f$. Using the Kelly result, $E_b\,{\rm Rate}[f] = D_1(b||m)$, we compute from Eq.~(\ref{Eq:RExpectedLogReturn})
\begin{equation}
E_b\,{\rm Rate}[f]-E_b\,{\rm Rate}[F]=\frac{R-1}{R}\,\Big(D_1(b||m)-D_{1/R}(b||m)\Big)\,.
\end{equation}
Using the fact that $D_\alpha$ is a nondecreasing function in $\alpha$ we can rewrite this as
\begin{equation}\label{Eq:final}
E_b\,{\rm Rate}[f]-E_b\,{\rm Rate}[F]=\frac{|R-1|}{R}\,\Big|D_1-D_{1/R}\Big|\,.
\end{equation}

Equations (\ref{Eq:RExpectedLogReturn}), (\ref{Eq:GeneralExpectedLogReturn}) and (\ref{Eq:final}) convert the abstract axiomatically-motivated measure of R\'enyi divergence into a much more intuitive measure of financial returns. This is the point where pure and simple mathematics turns into science about the real world, so we need to be extra careful: we need to understand the limitations of the resulting economic intuition before it can be used in practice.

\subsection{Reality check -- risk aversion of equity investors}\label{Sec:RiskAversionRange}

Using the rational part of Samuelson's critique as inspiration, we demand, as a matter of principle, that economic intuition can only be based on realistic investors. Failing that, we might not be able to relate correctly to the calculated experience, so the resulting financial intuition might be flawed.

Putting aside the information-theoretic definitions, we see that all our expressions for financial returns follow directly from the payoff elasticity equation~(\ref{Eq:PayoffElasticity}). We therefore require Eq.~(\ref{Eq:PayoffElasticity}) to be tested for its ability to explain the observed financial returns.

In this section, we review the observed data on equity returns, formulate the relevant tests and summarize their results. In terms of practical outcomes, we estimate the range of risk aversion for which the financial intuition is safe to use (the domain of Fig.~1B). The amount of technical detail imbedded in this section might seem out of proportion to the relatively simple topic of this paper. For this reason we defer as many technical arguments as we can to the supplementary materials paper~\cite{Soklakov_2014EqPuzzle}.

The type and the amount of available data depend greatly on a financial asset. Often we have some records of past performance (i.e. realized returns). For a popular established asset we might also have records of investors' past expectations (expected returns). Both types of returns are interesting for us (as they would be for any investor making a decision).

Some assets might also support a derivatives market. Price records from such markets effectively provide us with a history of market-implied distributions. Imagine, for instance, that in equation~(\ref{Eq:Price}) we knew ${\rm Price}[F]$ for any $F$. That would give us the distribution $m(x)$ which is a very important element in our theory ($m$ enters Eq.~(\ref{Eq:PayoffElasticity}) via $f$).

In what follows we focus on the famous S\&P 500 index and the relevant derivatives markets. The value of the index is proportional to the aggregated capitalization of 500 large companies listed on US stock exchanges. The total returns on the index (i.e. returns including dividends) reflect a fairly broad equity investment. The records of the index go back to its inception in 1926. The records of derivative prices and the investor-believed returns are much more recent by comparison (see below). Nevertheless, the S\&P 500 is an excellent benchmark of equity performance which is widely used in both industry and academia.

The above mentioned data contain a lot of nontrivial information which continues to challenge major economic theories. For instance, the entire class of consumption-based models have been struggling to explain the data for over 30 years. This fact is widely known in economics as the equity premium puzzle (see Ref.~\cite{Mehra_2008} and references therein).

In order to see how we can test Eq.~(\ref{Eq:PayoffElasticity}) we need to understand its place in the bigger picture. The equation describes a rational strategy. In other words, it provides a solution to a standard optimization with the standard form of expected utility~(\ref{Eq:ExpectedUtility}). However, unlike the consumption-based models in economics, the payoff elasticity equation~(\ref{Eq:PayoffElasticity}) does not claim to describe the entirety of human economic behaviour.

It is a key scientific fact that the human brain is (simultaneously) engaged in many strategies.\footnote{Even a single strategy can have multiple (simultaneous) implementations within the neocortex~\cite{HawkinsEtAl_2019}.} Each of these strategies has a goal. In this sense all individual strategies are rational by definition (with respect to their individual narrow goals). Equation~(\ref{Eq:PayoffElasticity}) can be very useful in describing such individual strategies.

The individual strategies are in constant competition with each other for limited resources. Even within a single person this competition produces a highly complex behaviour which, as far as we know, does not fit any simple model. Equation~(\ref{Eq:PayoffElasticity}), or indeed any single-goal rational framework, fails to describe a human person (let alone an economy).

In order to test Eq.~(\ref{Eq:PayoffElasticity}) we need to isolate a single strategy. Competition between strategies helps us to do that. Indeed, competition promotes strategies which make sense, strategies which justify themselves. In particular, a simple strategy which demonstrates realistic expectations that are regularly confirmed by actual performance has a good chance of becoming popular. The popularity of such strategies makes their effects measurable on a large scale.

Equity investment is a very simple and popular strategy. Specialized infrastructure in the form of stock exchanges and numerous trading firms support huge transaction volumes (currently approaching $10^8$ transactions per day globally). Equity investments are easy to liquidate. To sustain high levels of popularity for many decades (if not a century) the strategy of equity investment must make a lot of sense.

This gives us an opportunity to test Eq.~(\ref{Eq:PayoffElasticity}). Indeed, if the equation claims to describe a real investment product, it must apply to a simple equity investment. In the context of financial returns we identify three key types of tests: (i) understanding investor-expected returns, (ii) understanding realized returns and (iii) demonstrating consistency between the investor-expected and the realized returns.

Let us now progress to a more technical level and see how such tests can be performed. First let us examine the structure of Eq.~(\ref{Eq:PayoffElasticity}). We notice that Eq.~(\ref{Eq:PayoffElasticity}) involves four quantities: the payoff $F$, the investor's view $b$, the market-implied $m$ and the investor's risk aversion $R$. In the case of a simple equity investment the payoff structure, $F$, is known exactly. The data from derivatives also gives us $m$ (as explained above). Eliminating these quantities from Eq.~(\ref{Eq:PayoffElasticity}) leaves us with the connection between $b$ and $R$. Equivalently, we can speak of a family of investor-believed distributions parameterized by their risk aversion: $b\in\{b_R\}_R$. This observation is very useful for understanding the logic of the tests.

Since we know $F$, the reader should not be surprised that thinking in terms of $b_R$ might allow us to compute the expected rate of return as a function of $R$ (see Eq.~(\ref{Def:ExpectedLogReturn})). So, if we know what the investors are expecting, as we do in test (i), we can deduce the investor's risk aversion $R$.

In the supplementary materials paper~\cite{Soklakov_2014EqPuzzle} we complete the above logic with all the details. The independent quotes on the implied equity premium~\cite{Damodaran_2014} provide us with data on the investor-expected returns. The research reported in~\cite{Soklakov_2014EqPuzzle} was based on the most current data available at the time. This covered the period of time from September 2008 to April 2015 (monthly quotes at the beginning of each month, see Fig.1 of Ref.~\cite{Soklakov_2014EqPuzzle}). In terms of derivatives data (which we need for $m$) this timeline is well within the recent history retained by most investment banks. For the readers who have no license to standard commercial data on derivatives Ref.~\cite{Soklakov_2014EqPuzzle} provides a simplified version of the calculations (which can also serve as a ball-park stability check for the main calculations).

We found that the values of $R$ corresponding to the investor-expected returns were mostly in the range between 1 and 2.5 (see Fig.~2 of~\cite{Soklakov_2014EqPuzzle}). By the standards adopted in the literature on the equity premium puzzle~\cite{Mehra_2008} these values are well within the expected norm. Certainly, looking at the original paper by Mehra and Prescott~\cite{MehraPrescott_1985}, we see that no puzzle would ever have been reported if these values of risk aversion had been implied by a consumption-based model. This completes the test of Eq.~(\ref{Eq:PayoffElasticity}) for its consistency with the observed investor-expected returns (test-(i) in the above enumeration).

In the second type of tests (on realized returns) we do not have $b$. Instead we have an historical distribution of returns. We know, however, that the investor who happened to have the correct view will find the long-term realized returns approaching their expectations. Mathematically, this can be seen by substituting $p=b$ into Eq.~(\ref{Eq:GeneralExpectedLogReturn}) and comparing the result to Eq.~(\ref{Eq:RExpectedLogReturn}). This is in fact how the investors' expectations materialize in practice (see section 2.2 of~\cite{Soklakov_2014EqPuzzle} for detailed explanations).

Thus, by matching $b_R$ to the historical distribution, we can once again estimate the values of $R$. This time, however, the numerical values of risk aversion are implied by the historical (i.e. realized) distribution.

Accurate representation of a distribution by a sample requires a lot of data. The data must include the derivatives market (because we need $m$). We managed to find daily records on both the equity returns and, crucially, the derivatives market dating back to 17th May 2000. The last day in the sample is 27th April 2015. We found the range of $R$ explaining the realized returns was between 0.5 and 3 (see Fig.~4 of the supplementary materials paper~\cite{Soklakov_2014EqPuzzle}). Just like we saw in the case of investor-expected returns (test-(i)), these values of risk aversion are well within the expectations.

For test-(iii) we need to compare our findings regarding the investor-expected and the realized returns. We see that the corresponding ranges of $R$ overlap showing good agreement. More accurately, we notice the range of $0.5 \lesssim R \lesssim 3$ explaining the realized returns is slightly wider than the corresponding range explaining the investor-expected returns, $1 \lesssim R \lesssim 2.5$. This is indeed what we should expect.

We expect our test of the realized returns to overestimate the range for $R$ because real investors do make mistakes in their forecasts. For example, we can find historical periods with very low (or even negative) realized returns. Rational investors would accept such returns only if they had very low levels of risk aversion. In reality, however, the investors enter such periods unawares so their actual $R$ might not be as low as suggested by the data. By trying to explain all of the data (as was done in~\cite{Soklakov_2014EqPuzzle}) we assume that the investor foresaw the realized distribution of returns and invested in full knowledge of that distribution. This overestimates the range of $R$ that is necessary to explain the observations.

Consequently, we use the values of $R$ between 1 and 2.5 as confirmed and think of the wider range (between 0.5 and 3) as a possible overestimation.

Future investigations using Eq.~(\ref{Eq:PayoffElasticity}) for different types of investors (not necessarily equity) and considering broader time periods will deepen our understanding of the possible ranges for $R$. As we can see from the above formulae, this in turn will provide better intuition over the R\'enyi parameter~$\alpha$.

\subsection{Further reality check -- the neuroscience of decision-making}\label{Sec:Neuroscience}

For better or worse the concept of financial returns has been very influential in human decision making. Indeed, even in the narrow field of finance, investment returns are by no means the only characteristic that one could compute about a business and yet it is certainly a very popular (if not the most popular) quantity that is always discussed during any investment decision process. Even the critics of the financial industry acknowledge the use of financial returns as a prime decision quantity (albeit as a manifestation of greed rather than a rational procedure).

Could it be that our brains naturally operate with quantities akin to financial returns when implementing at least some types of decision making? Below we give a cautiously positive answer to this question.

We turn our attention to the neurophysiological experiments reported in~\cite{YangShadlen_2007,KiraEtAl_2015}. In these experiments rhesus monkeys were presented with a binary decision task under uncertainty. Hoping for a reward, the monkeys communicated their decision by making an eye movement to either a green or a red target.

The monkeys had to base their decisions on a sequence of (visually presented) abstract shapes which influenced the reward probabilities in a certain well-controlled manner. Some shapes supported the green choice while others suggested the red target. The statistical meaning of each individual shape was fixed and the monkeys were given enough training to learn the general direction (red or green) as well as the shapes' relative importance.\\ \\

During the decision process the electrical activity of individual neurons in the lateral intraparietal area was measured. This area is rich in neurons which are believed to be involved in planning eye movements~\cite{GnadtAndersen_1988, PlattGlimcher_1997, ColbyGoldberg_1999, IptaEtAl_2006}. The electrical activity of such neurons reliably predicts whether the target of the planned movement lies within or outside a fixed small region of the visual field (the response field of a neuron). It was found experimentally that the firing rate of such neurons is proportional to the log-likelihood ratio~\cite{YangShadlen_2007}:
\begin{equation}\label{Eq:FiringRate}
{\rm FiringRate} \propto \ln\frac{p(s_1,\dots,s_N| {\rm\, reward\ \lq\lq in")}}{p(s_1,\dots,s_N| {\rm\, reward\ \lq\lq out"})}\,,
\end{equation}
where $s_1,\dots,s_N$ is the sequence of shapes shown to the monkey, $p(s_1,\dots,s_N| {\rm\, reward\ \lq\lq in"})$ is the conditional probability of the sequence given the rewarded target appears is in the neuron's response field and $p(s_1,\dots,s_N| {\rm\, reward\ \lq\lq out"})$ is the conditional probability of the sequence given the rewarded target lies outside the neuron's response field.

The gradual neurophysiological accumulation of probabilistic evidence suggested by the above expression has been explicitly tested~\cite{KiraEtAl_2015}. The same experiments revealed that, given the choice, the monkeys would voluntarily terminate the process of evidence accumulation and express their decision once the firing rate~(\ref{Eq:FiringRate}) reaches a certain critical value.

Let us now examine what the above experiments say in the context of the payoff elasticity equation~(\ref{Eq:PayoffElasticity}). Using the notation of Eq.~(\ref{Eq:FiringRate}), the binary outcome for the location of the reward in or out of the response field of the measured neuron corresponds to a binary random variable $X=\{{\rm\, reward\ \lq\lq in"}, {\rm\, reward\ \lq\lq out"}\}$. In this notation
\begin{equation}
d\ln F(x) =\ln F({\rm\, reward\ \lq\lq in"})-\ln F({\rm\, reward\ \lq\lq out"})=\ln\frac{F({\rm\, reward\ \lq\lq in"})}{F({\rm\, reward\ \lq\lq out"})}\,.
\end{equation}
Writing the analogous equation for $d\ln f(x)$ and using Eq.~(\ref{Eq:PayoffElasticity}) we compute
\begin{equation}\label{Eq:BinaryPayoffElasticity}
\ln\frac{F({\rm\, reward\ \lq\lq in"})}{F({\rm\, reward\ \lq\lq out"})}=\frac{1}{R}\,\ln\frac{f({\rm\, reward\ \lq\lq in"})}{f({\rm\, reward\ \lq\lq out"})}\,.
\end{equation}
Now let us recall the interpretation of $f$ as the likelihood function~\cite{Soklakov_2011}. This comes from understanding the equation
\begin{equation}
b(x)=f(x)\,m(x)
\end{equation}
as Bayes' theorem
\begin{equation}\label{Eq:Bayes}
p(x|s_1,\dots,s_N)=\frac{p(s_1,\dots,s_N|x)}{p(s_1,\dots,s_N)}\,p(x)\,.
\end{equation}
The market-implied distribution, $m(x)$, is perceived by the investor as prices. In the above experiments, for example, there was no a priori price difference between the red and the green targets. In other words, the market was flat: $m({\rm\, reward\ \lq\lq in"})=m({\rm\, reward\ \lq\lq out"})$. With no other data, i.e. before any further learning took place, $m(x)$ defines the prior distribution $p(x)\equiv m(x)$. The investor-believed distribution reflects the investor's final state of knowledge, i.e. it is the posterior $b(x)\equiv p(x|s_1,\dots,s_N)$. It follows that $f$ is the likelihood
\begin{equation}
f(x)=\frac{p(s_1,\dots,s_N|x)}{p(s_1,\dots,s_N)}\,.
\end{equation}
Using this fact we can rewrite Eq.~(\ref{Eq:BinaryPayoffElasticity}) as
\begin{equation}\label{Eq:BinaryPayoffElasticity2}
\ln\frac{F({\rm\, reward\ \lq\lq in"})}{F({\rm\, reward\ \lq\lq out"})}=\frac{1}{R}\,\ln\frac{p(s_1,\dots,s_N|{\rm\, reward\ \lq\lq in"})}{p(s_1,\dots,s_N|{\rm\, reward\ \lq\lq out"})}\,.
\end{equation}

By comparing this equation with the experimentally observed relationship~(\ref{Eq:FiringRate}), we derive
\begin{equation}\label{Eq:ReturnPropFiringRate}
\ln\frac{F({\rm\, reward\ \lq\lq in"})}{F({\rm\, reward\ \lq\lq out"})}\propto\,{\rm FiringRate}\,,
\end{equation}
where the proportionality coefficient may depend on risk aversion $R$.

On the right-hand side of proportionality relation~(\ref{Eq:ReturnPropFiringRate}) we have a quantity which is directly involved in decision making -- the firing rate of the relevant neurons. The quantity on the left-hand side is about the optimal investment payoff $F$; it is the difference of log-returns between the two possible outcomes inside the optimal investment.

Although we should be careful not to overstate the importance of (\ref{Eq:ReturnPropFiringRate}), it does provide us with some evidence that the concept of financial returns on optimized investments is not at all alien to our brains. Sometimes it can even be measured directly from individual neurons.

\subsection{Further mathematical generalizations}\label{Sec:MathGeneralizations}
The connection between divergence measures and investment behaviors turns out to be rather general. Consider for instance a general investor whose relative risk aversion is a function of the payoff, i.e. $R=R(F)$. Equation~(\ref{Eq:PayoffElasticity}) shows that the corresponding optimal payoff is a function of the growth-optimizing payoff $F(\cdot)=\tilde{F}(f(\cdot))=\tilde{F}(b(\cdot)/m(\cdot))$, so the expected logarithmic rate of return takes the form
\begin{equation}
E_b\,{\rm Rate}[F]=\int b(x)\,\phi\left(\frac{m(x)}{b(x)}\right)\,dx
\end{equation}
for some function $\phi$. This is exactly the form of the $\phi$-divergence measures which were introduced and studied independently in Refs.~\cite{Csiszar_1963}, \cite{Morimoto_1963} and \cite{AliSilvey_1966}. The financial interpretation thus naturally extends to a very broad class of divergence measures. Different divergence measures appear connected to different investment behaviors.

Some readers may wonder whether we could consider the concept of utility instead of investment returns. In the most general case, we would expect the financial intuition to be lost due to the affine freedom in the definition of utility. Indeed, one can multiply utility by a constant or add an arbitrary number with no effect on the optimal choices. The concept of utility effectively ignores the two most simple and intuitive mathematical operations. We expect this to be detrimental to intuition. However, at least in some cases, there are certain mathematical similarities (cf.~\cite{BleulerEtAl_2019,BleulerEtAl_2020}). Such similarities may be useful for extending our financially-intuitive approach to the study of utility.\\



\end{document}